# A Toy Model for Magnetized Neutrino-Dominated Accretion Flows

Lei WeiHua, Wang DingXiong[†], Zhang Lei, Gan ZhaoMing & Zou YuanChuan

School of Physics, Huazhong University of Science and Technology, Wuhan 430074, China

**In this paper, we present a simplified model for magnetized neutrino-dominated accretion flow (NDAF) in which effect of black hole (BH) spin is taken into account by adopting a set of relativistic correction factor, and the magnetic field is parameterized as $\beta$, the ratio of the magnetic pressure to the total pressure. It is found that the disc properties are sensitive to the values of the BH spin and $\beta$, and more energy can be extracted from NDAF for the faster spin and lower $\beta$.**

The nature of the central engine of gamma-ray bursts (GRBs) remains unclear, and the favoured models invoke binary merger or collapse of compact objects. These models lead to the formation of a transient hot and dense accretion torus/disc around a black hole (BH) of a few solar masses. If an accretion disc is cooled mainly via the neutrino loss, it is referred to as neutrino-dominated accretion flow (NDAF)[1].

NDAF has been extensively discussed by many authors [2-5]. However, the effects of the magnetic fields are usually neglected in NDAF due to the intrinsic complexity.

Magnetic fields could play an important role in the central engine in some aspects [10], e.g., a bold gamma-ray polarization [6], a higher magnetic energy density required by the reverse shock [7,8], and the energy argument for X-ray flares [9]. These considerations stimulate us to discuss a model of magnetized NDAF.

Recently, Chen & Beloborodov [4], Gu et al. [5] and Shibata et al. [11] argued that the general relativistic (GR) effects are important for NDAF, and we introduce GR correction factors in our model.

This paper is organized as follows. In Sect. 1 we describe the magnetized NDAF in the frame of a relativistic thin disc of steady state, and the effects of MHD stress are described. The main equations are based on refs. [2] and [3], but corrected with GR factors. We solve the set of equations for the solutions of NDAF in Sect. 2, and compute disc temperature, density and neutrino luminosities. The effects of the free parameters are also discussed. Finally, we summarize our results and discuss some related issues in Sect. 3.

## 1. Magnetized Neutrino-Dominated Accretion Flows

As is well known, the magneto-rotational instability (MRI) plays an important role in angular momentum transportation of accretion disc [11, 12]. Currently, it is widely believed that the viscosity in accretion flows is effectively determined by the turbulent motion of fluid, being related to the magnetic stress. Therefore, we consider only the magnetic viscosity in this model.

The magnetic viscous shear $t_{r\varphi}$ can be expressed as

$$t_{r\varphi} = -\frac{B_r B_\varphi}{4\pi}, \quad (1)$$

where $B_\varphi$ is given by

$$\frac{B_\varphi^2}{8\pi} = \beta P. \quad (2)$$

The parameter $\beta$ is the ratio of the magnetic pressure to the total pressure.

A roughly steady state is reached when the growing rate of $B_\varphi$ generated by differential rotation of the radial field is equal to its loss rate due to buoyancy effect, and $B_\varphi$ can be estimated as [13]

$$B_\varphi \approx \left[-B_r \frac{d\Omega_{disk}}{d\ln r} H\right]^{1/2} (4\pi\rho)^{1/4} \qquad (3)$$

MHD simulations of differentially-rotating discs indicate that $B_\varphi$ is the largest component of the magnetic field as shown in shearing box simulations (e.g., [14]) and also in global GRMHD simulations. It is easy to check that this requirement is satisfied in eqn.(3).

A plausible origin of $B_r$ is the pre-coalescence magnetic fields of the neutron stars. Values as large as $\sim 10^{13}G$ are observed for some radio pulsars.

The relativistic correction factors for a thin accretion disc around a Kerr BH have been given by [15],

$$A = 1 - \frac{2GM}{c^2 r} + (\frac{GMa_*}{c^2 r})^2, \qquad (4)$$

$$B = 1 - \frac{3GM}{c^2 r} + 2a_*(\frac{GM}{c^2 r})^{3/2}, \qquad (5)$$

$$C = 1 - 4a_*(\frac{GM}{c^2 r})^{3/2} + 3(\frac{GMa_*}{c^2 r})^2, \qquad (6)$$

$$D = \int_{r_{ms}}^{r} \frac{\frac{x^2 c^4}{8G^2} - \frac{3xMc^2}{4G} + \sqrt{\frac{a_*^2 M^3 c^2 x}{G}} - \frac{3a_*^2 M^2}{8}}{\frac{\sqrt{rx}}{4}(\frac{x^2 c^4}{G^2} - \frac{3xMc^2}{G} + 2\sqrt{\frac{a_*^2 M^3 c^2 x}{G}})} dx. \qquad (7)$$

The equation of the conservation of mass remains valid, while hydrostatic equilibrium in the vertical direction leads to a corrected expression for the half thickness of the disc [15, 16]:

$$H \simeq \sqrt{\frac{Pr^3}{\rho GM}} \sqrt{\frac{B}{C}}. \qquad (8)$$

The basic equations of magnetized NDAF are given as follows.

1) The continuity equation:

$$\dot{M} = -2\pi r v_r \Sigma. \qquad (9)$$

2) The total pressure consists of five terms, radiation pressure, gas pressure, degeneracy pressure, neutrino pressure and magnetic pressure as follows,

$$P = \frac{11}{12}aT^4 + \frac{\rho kT}{m_p}(\frac{1+3X_{nuc}}{4}) + \frac{2\pi hc}{3}(\frac{3}{8\pi m_p})^{4/3}(\frac{\rho}{\mu_e})^{4/3} + \frac{u_\nu}{3} + P_{mag}, \qquad (10)$$

where $P_{mag} = B_\varphi^2/8\pi$ is the magnetic pressure contributed by the tangled magnetic field in the disc, and $u_\nu$ is the neutrino energy density. The detailed expression for $u_\nu$ is defined in ref. [3].

3) Combining the conservation of the angular momentum with eqn.(9), we have

$$\frac{d}{dr}(\dot{M}l) = -\frac{d}{dr}(4\pi r^2 t_{r\varphi} H), \qquad (11)$$

where $l$ is the specific angular momentum of the accreting gas. Vanishing of $t_{r\varphi}$ at the last stable orbit of the thin disc leads to

$$\dot{M}r^2 \sqrt{\frac{GM}{r^3}} \frac{D}{A} = -4\pi r^2 t_{r\varphi} H = r^2 H(B_r B_\varphi). \qquad (12)$$

4) The equation for the energy balance is

$$Q^+ = Q^-, \qquad (13)$$

where $Q^+ = Q_{vis}$ represents the viscous dissipation, and $Q^- = Q_\nu + Q_{photo} + Q_{adv}$ is the total cooling rate due to neutrino losses $Q_\nu$, photodisintegration $Q_{photo}$ and advection $Q_{adv}$. The detailed expressions for $Q_{photo}$, $Q_{adv}$ and the bridging formula for $Q_\nu$ are given in [3]. The heating rate $Q_{vis}$ is expressed as

$$Q_{vis} = \frac{3GM\dot{M}}{8\pi r^3} \frac{D}{B}. \qquad (14)$$

We solve eqns.(1)-(14) numerically for the solutions of disc temperature $T$ and density $\rho$ versus disc radius with the given $a_*$, $\beta$ and $\dot{m}_{acc}$, where $\dot{m}_{acc}$ is the accretion rate in units of $M_\odot s^{-1}$. The cooling term arising from the photodisintegration is not included, because it is much less than the neutrino cooling rate in the inner disc [17]. It is emphasized that this model is only appropriate for extremely super-Eddington accretion.

## 2. Numerical Results

There are four free parameters in our model, i.e. the BH mass $m$ (in units of $M_\odot$) and spin $a_*$, $\beta$ and $\dot{m}_{acc}$,

where the spin is related to the GR effects.

We are interested primarily in the properties of the inner accretion flow, where the neutrino process is important. As argued in [1]-[3], the flows are fully advection-dominated for $r > 100 r_g$, since neutrino cooling is not important and photons are completely trapped. Thus the discussion is focused in the region from $r_{ms}$ to $r_{max} = 100 r_g$.

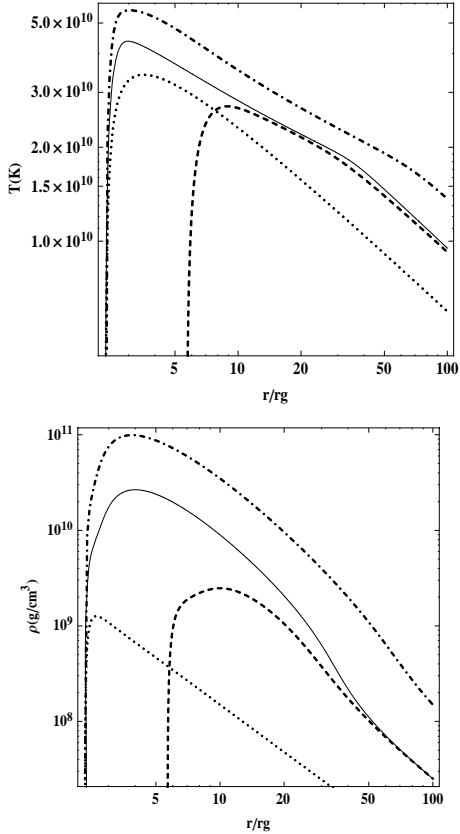

**Figure 1** The profiles of disc temperature $T$ and density $\rho$ versus disc radius for $\dot{m}_{acc} = 0.1$ in upper and lower panels, respectively. The parameters ($a_*$, $\beta$, $m$) are (0.9, 0.1, 7), (0.1, 0.1, 7), (0.9, 0.3, 7) and (0.9, 0.1, 3) in solid, dashed, dotted and dot-dashed lines respectively.

The profile of disc temperature $T$ and density $\rho$ versus disc radius for different $a_*$, $\beta$ and $m$ are shown in Figure 1.

First, we find that the BH spin has little effect at larger radius, but has significant effect in the inner region. This result can be understood as follows. As the spin becomes faster, e.g., $a_* = 0.9$, the disc extends to a smaller radius and has a wider inner region for releasing accretion energy.

Second, we note that the result with $\beta = 0.1$ is significantly different from that with $\beta = 0.3$, displaying that lower $\beta$ has higher temperature and density. This result implies that the region of neutrino emission extends to larger radius in the disc with a lower beta.

To understand the above result, we use the $\alpha$-prescription for magnetic viscous, i.e.,

$$\frac{B_r B_\varphi}{4\pi} = \alpha P \sqrt{\frac{A}{BC}}, \quad (15)$$

Combining eqns. (15) with (2), we have the viscous parameter $\alpha \leq 2\beta$, which varies with disc radius instead of a constant in our model. Therefore, the model with lower $\alpha$ has higher temperature and density, which is consistent with the results given in refs. [1] and [4].

Finally, we find that the disc is hotter and denser for the lower BH mass, which can be well understood by eqn. (14).

It is more interesting to investigate the neutrino luminosity from the accretion flow, which is defined as

$$L_\nu = 4\pi \int_{r_{ms}}^{r_{out}} Q_\nu r dr. \quad (16)$$

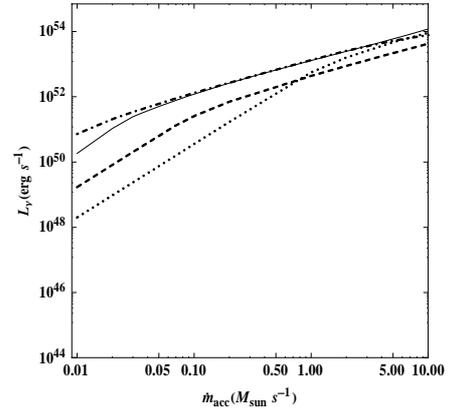

**Figure 2** The total neutrino luminosity $L_\nu$ as a function of $\dot{m}_{acc}$. The solid, dashed, dotted and dot-dashed lines have the same meanings as given in Figure 1.

Total neutrino luminosity $L_\nu$ versus $\dot{m}_{acc}$ is shown in Figure 2 for different $a_*$, $\beta$ and $m$. First, we find that the effect of the BH spin is significant for all accretion rates. This result can be easily understood based on the above discussion. Second, the effect of $\beta$ is important only for low accretion rates. And the neutrino optical depth becomes important for lower $\beta$ and larger accretion rate (see Figure 3). We also find that $L_\nu$ is independent of BH mass for large accretion rate. This is because most of the viscous dissipation converts to the neutrino emission at large accretion rate.

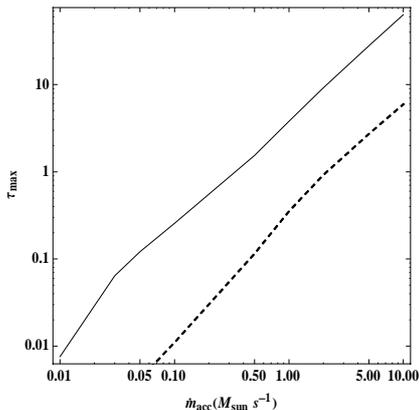

**Figure 3** The maximum neutrino optical depth as a function of $\dot{m}_{acc}$ for $a_* = 0.9$ and $m = 7$. The results for $\beta = 0.1$ and $\beta = 0.3$ are shown in solid and dashed lines, respectively.

## 3. Conclusion and discussion

In this paper, we investigate the effects of BH mass, spin and $\beta$ on NDAF in a simplified model. It is found that more energy can be extracted from NDAF for fast BH spin, low $\beta$.

A remarkable feature of this model is that only four free parameters, $a_*$, $\beta$, $m$ and $\dot{m}_{acc}$, are involved rather introducing an unknown viscous parameter $\alpha$. In addition, the Blandford-Znajek process [18], Blandford-Payne process [19] and magnetic coupling process [20, 21] can be incorporated with NDAF based on this model.

The large-scale magnetic field of $B \sim 10^{15} G$ required in these processes could be realized, provided that the Parker instability or buoyant rise convert $B_\varphi$ to $B_z$ with reasonable efficiency [13]. These attempts may be helpful to explore the mechanism for X-ray flares and shallow decays observed in the early afterglows of some GRBs by Swift.

**Acknowledgments.** Supported by National Natural Science Foundation of China under Grants 10873005, 10847127 and 10703002, the Research Fund for the Doctoral Program of Higher Education under grant 200804870050 and National Basic Research Program of China - 973 Program 2009CB824800